\documentclass[12pt,preprint,apj]{emulateapj}

\newcommand{\kms}{{\rm km~s^{-1}}}

\newcommand{\rvir}{R_{200}}
\newcommand{\sfropt}{\rm SFR_{SDSS}}

\shorttitle{MIR properties of A2199 Supercluster}
\shortauthors{Lee et al.}

\begin{document}

\title{Galaxy Evolution in the Mid-infrared Green Valley: a Case of the A2199 Supercluster}
 
\author{Gwang-Ho Lee$^{1}$, Ho Seong Hwang$^{2,3}$, Myung Gyoon  Lee$^{1}$, 
Jongwan Ko$^{4}$, Jubee Sohn$^{1}$, Hyunjin Shim$^{5}$, 
and Antonaldo Diaferio$^{6,7}$}
\affil{$^{1}$Department of Physics and Astronomy, Seoul National University , 1 Gwanak-ro, Gwanak-gu, Seoul 151-742, Republic of Korea}
\email{ghlee@astro.snu.ac.kr}
\email{mglee@astro.snu.ac.kr}
\affil{$^{2}$Smithsonian Astrophysical Observatory, 60 Garden Street, Cambridge, MA 02138, USA}
\affil{$^{3}$School of Physics, Korea Institute for Advanced Study,
   85 Hoegiro, Dongdaemun-gu, Seoul 130-722, Republic of Korea}
\affil{$^{4}$ Korea Astronomy and Space Science Institute, Daejeon 305-348, Republic of Korea}   
\affil{$^{5}$Department of Earth Science Education, Kyungpook National University, Republic of Korea}
\affil{$^{6}$Dipartimento di Fisica, Universit\`a di Torino, Torino, Italy}
\affil{$^{7}$Istituto Nazionale di Fisica Nucleare (INFN), Sezione di Torino, Torino, Italy}




\begin{abstract}
We study the mid-infrared (MIR) properties of the galaxies
  in the A2199 supercluster at $z=0.03$
  to understand the star formation activity of galaxy groups and clusters
  in the supercluster environment.
Using the \textit{Wide-field Infrared Survey Explorer} data,
  we find no dependence
  of mass-normalized integrated SFRs of galaxy groups/clusters
  on their virial masses. 
We classify the supercluster galaxies
  into three classes in the MIR color-luminosity diagram: 
  MIR blue cloud (massive, quiescent and mostly early-type), 
  MIR star-forming sequence (mostly late-type), 
  and MIR green valley galaxies. 
 These MIR green valley galaxies are distinguishable from the optical green valley galaxies, 
  in the sense that they belong to the optical red sequence.
We find that the fraction of each MIR class does not depend on 
  virial mass of each group/cluster. 
We compare the cumulative distributions of 
  surface galaxy number density and 
  cluster/group-centric distance for the three MIR classes.
MIR green valley galaxies show the distribution between
  MIR blue cloud and MIR SF sequence galaxies.
However, if we fix galaxy morphology,
  early- and late-type MIR green valley galaxies
  show different distributions.
Our results suggest a possible evolutionary scenario of these galaxies:
1) Late-type MIR SF sequence galaxies $\rightarrow$
2) Late-type MIR green valley galaxies $\rightarrow$
3) Early-type MIR green valley galaxies $\rightarrow$
4) Early-type MIR blue cloud galaxies.
In this sequence, star formation of galaxies is quenched before the galaxies enter 
  the MIR green valley, and then morphological transformation occurs in the MIR green valley. 
\end{abstract}


\section{Introduction}

Color bimodality of galaxies appears in several wavelengths including
  the optical \citep{strateva+01,baldry+04,choi+07}, 
  ultraviolet \citep{wyder+07,brammer+09}, 
  and mid-infrared (MIR) bands \citep{johnson+07,walker+13}.
In optical color-magnitude diagrams
  galaxies are often separated into two classes:
  a red sequence and a blue cloud.
The evolution between the two populations is
  one of the key issues in the recent study
  of galaxy formation and evolution \citep{faber+07,tinker+13}.

Galaxies in a region between the red sequence and the blue cloud
  (i.e., green valley)
  seem to be an important population to understand
  this evolution.
They might be in a transition phase
  from the blue cloud toward the red sequence;
  the transition would occur during a short timescale less than 1 Gyr 
  \citep{springel+05,faber+07,pan+13}.  
Recently, \citet{schawinski+14} suggested that 
  the green valley galaxies are not in a single transitional state
  from the blue cloud toward the red sequence.
They found that 
  early- and late-type galaxies in the green valley of optical color-mass diagrams
  have significantly different ultraviolet$-$optical colors.
They pointed out that this color difference could be explained 
  by different quenching timescales of star formation;
  quenching timescale for early-type galaxies ($<250$ Myr) is much shorter than 
  that for late-type galaxies ($>1$ Gyr).

Galaxies can also be divided into several classes based on 
  MIR colors such as 
  AKARI $N3~(3.2~\micron)-S11~(10.4~\micron)$ \citep{shim+11,ko+12} 
   and \textit{Spitzer} IRAC $3.5~\micron-8~\micron$ 
   \citep{johnson+07,walker+13}. 
These MIR colors are useful indicators of 
  specific star formation rates \citep{duc+02,chung+11}
  and of mean stellar ages \citep{piovan+03,temi+05,ko+09}.
For example, 
  \citet[][hereafter H12]{hwang+12} studied the galaxies of the A2199 supercluster
  in the $[3.4]-[12]$ color versus 
  12 $\micron$ luminosity diagram using the 
  \textit{Wide-field Infrared Survey Explorer} 
  \citep[\textit{WISE},][]{wright+10} data.
They found that early- and late-type galaxies 
  are well separated into two groups;
  late-type galaxies with MIR red colors form 
  the ``MIR star-forming (SF) sequence'', 
  but early-type galaxies mainly form the ``MIR blue cloud''.

As in the optical color-magnitude diagrams,
  galaxies with intermediate MIR colors seem to be
  transition populations (e.g., \citealt{shim+11,ko+12,ko+13}).
Recently, \citet{alatalo+14} defined the infrared transition zone 
  in the \textit{WISE} $[4.6]-[12]$ color space, and suggested that 
  the galaxies in this zone are in the late stages of transition
  across the optical green valley.
However, how this transition differs
  depending on galaxy morphology is not well understood.
This is the goal of this study;
  we define ``MIR green valley'' galaxies 
  in the MIR color-luminosity diagram, and study their properties
  in connection to other populations.

On the other hand,
  galaxy properties including the star formation activity  
  and the galaxy morphology are strongly affected by environment \citep{park+07,bm09}.
  Galaxy clusters are an ideal place to study this environmental dependence of 
  galaxy properties, and several physical mechanisms in the cluster environment 
  have been proposed to explain the dependence:
  ram pressure stripping \citep{gunn+72}, thermal evaporation \citep{cowie+77},
  strangulation \citep{larson+80}, galaxy harassment \citep{moore+96}, 
  starvation \citep{bekki+02}, and cumulative galaxy-galaxy hydrodynamic/gravitational 
  interactions \citep{park+09}.

\begin{deluxetable*}{cccccccccc}
\tablecolumns{10}
\tablewidth{0pc}
\tablecaption{SDSS-related physical parameters of member galaxies}
\tablehead{
\colhead{ID} & \colhead{SDSS ObjID (DR7)} & \colhead{R.A. (J2000)} & \colhead{Decl. (J2000)} & \colhead{$z$} &
\colhead{Morph$^{a}$} & \colhead{${\rm log}(M_{\rm star}/M_{\odot})$} & \colhead{$u-r^{b}$} & \colhead{${\rm log SFR_{SDSS}}$} & \colhead{$D_n4000$} }
\startdata
 1 & 587725992501969046 & 16:46:57.61 & 41:56:22.7 & 0.032715 & 2  &  9.703  &  1.075  &  1.288  &  1.257 \\
 2 & 587725992503017882 & 16:53:52.93 & 39:49:46.7 & 0.033889 & 1  &  9.816  &  2.543  &  0.035  &  1.655 \\
 3 & 587725992503017689 & 16:53:56.32 & 39:48:45.4 & 0.033287 & 1  & 10.794  &  2.876  &  0.049  &  1.961 \\
 4 & 587725992502820896 & 16:52:29.10 & 40:18:43.0 & 0.029247 & 1  & 10.765  &  2.857  &  0.140  &  2.006 \\
 5 & 587725993036939572 & 16:36:03.51 & 45:44:47.6 & 0.030912 & 2  & 10.524  &  2.027  &  1.188  &  1.750
\enddata
\label{cat_sdss}
\tablecomments{$^{a}$ Morphology classification: 1- Early types (E/S0), 2- Late types (S/Irr). $^{b}$ Extinction-corrected $u-r$}
\end{deluxetable*}

\begin{deluxetable*}{cccccccc}
\tablecolumns{8}
\tablewidth{0pc}
\tablecaption{\textit{WISE}-related physical parameters of member galaxies}
\tablehead{
\colhead{ID} & \colhead{\textit{WISE} ID} & \colhead{$[3.4]-[12]$} & \colhead{${\rm log} (L_{12\micron}/L_{\odot})$} & \colhead{(S/N)$_{12\micron}$} & \colhead{MIRcl$^{a}$} & \colhead{logSFR$_{\rm WISE}$} & \colhead{(S/N)$_{22\micron}$} }
\startdata
1 & J164657.63+415622.9  &  1.389  &  8.564 & 21.60 & 3 &  0.683 &  6.30 \\
2 & J165352.94+394946.7  & $-0.600$  &  7.680 &  1.40 & 0 &  0.255 & $-1.70$ \\
3 & J165356.33+394845.4  & $-1.879$  &  8.106 &  8.40 & 1 &  0.404 &  3.30 \\
4 & J165229.09+401843.0  & $-1.934$  &  8.096 & 11.40 & 1 &  0.311 &  1.20 \\
5 & J163603.52+454447.8  &  0.406  &  8.753 & 34.20 & 3 &  0.739 &  7.70
\enddata
\label{cat_wise}
\tablecomments{$^{a}$ MIR classification: 0- (S/N)$_{12\micron}<3$, 1- MIR blue cloud galaxies, 
2- MIR green valley galaxies, 3- MIR star-forming sequence galaxies.}
\end{deluxetable*}

Some cluster galaxies might have experienced 
  the environmental effects even before they enter the cluster regions. 
  For example, \citet{zabludoff98} suggested that massive early-type galaxies 
  are formed by galaxy-galaxy mergers in group-scale environment, 
  and then fall into clusters later. 
  The galaxy interactions and mergers should be very frequent and effective 
  because of low velocity dispersion of galaxies in the group environment
  \citep{hickson+92,sohn+13}.
  This ``pre-processing'' in groups is supported 
  by numerical simulations \citep{fujita04} and by observations 
  \citep{koyama+11,lemaux+12,hess+13,mahajan13}.
  Therefore, it is important to study the galaxies in galaxy groups 
  to better understand the environment effects on galaxy properties.
  
Because superclusters of galaxies contain
  several clusters and groups,
  they can be excellent laboratories 
  for studying the environmental dependence of galaxy properties
  in a full range of environment from cluster to field 
  (e.g., \citealt{haines+11, biviano+11}).
In this study, we use a multiwavelength catalog of the galaxies
  in the A2199 supercluster at $z=0.03$
  to study the MIR properties of galaxies and their 
  environmental dependence.
The main goal is to examine how the galaxies in the 
  MIR green valley differ from other populations
  as a function of environment.
    
\S \ref{data} describes the observational data we use.
We derive physical parameters of the groups and clusters 
  in the A2199 supercluster in $\S$\ref{basic}. 
We investigate the dependence of 
  mass-normalized integrated star formation rates (SFRs)
  of galaxy groups and clusters
  on their virial masses in $\S$\ref{isfr}.
We classify the galaxies in the MIR color-luminosity diagram,
  and study the environmental dependence of their MIR properties
  in $\S$\ref{mirc}.
We discuss our results in $\S$\ref{discuss}, and
  conclude in $\S$\ref{conclusions}.
Throughout, we adopt flat $\Lambda$CDM cosmological parameters:
  $H_{0}=70~\kms~{\rm Mpc^{-1}}$, $\Omega_{\Lambda}=0.7$, and $\Omega_{m}=0.3$.

\section{Data}\label{data}

We used the galaxy catalog of the A2199 supercluster given by H12.  
H12 compiled the multi-wavelength data of the galaxies at $m_r<17.77$ 
  from the Sloan Digital Sky Survey Data Release 7 \citep{abazajian+09} 
  and the \textit{WISE} data.
The \textit{WISE} provides
  all-sky survey data in four MIR bands (3.4, 4.6, 12, and 22 $\micron$)
  with a much better sensitivity than previous infrared surveyors.
Thus, the data cover the entire region 
  of the A2199 supercluster at $z=0.03$ 
  ($12\degr\times12\degr\simeq27~{\rm Mpc}\times27~{\rm Mpc}$).
  
H12 selected member galaxies using the caustic method
  \citep{diaferio97,diaferio99,serra+11}.
They also compiled redshift data for the galaxies 
  fainter than $m_r=17.77$ from other 
  large spectroscopic surveys \citep{rines+02,rines+08}.
However, these surveys cover only the central region ($R<50$ arcmin) 
  of the A2199 supercluster.
Therefore, we did not use these faint galaxies, 
  but used the galaxies at $m_{r}<17.77$ 
  to have a galaxy sample with uniform depth
  in the entire supercluster region. 
The final sample contains 1529 member galaxies including
  559 early-type (E/S0) and 970 late-type galaxies (S/Irr).

Galaxy morphology information is mainly from 
  the Korea Institute for Advanced Study (KIAS) 
  DR7 value-added galaxy catalog \citep{choi+10}.
  The morphology information is based on the automatic classification 
  by \citet{park+05}, which uses $u-r$ color, $g-i$ color gradient, 
  and $i$−band concentration index. 
  The completeness and reliability of this method are $>90\%$.
  H12 conducted an additional visual inspection to improve the classification results 
  and to classify the galaxies not included in the KIAS DR7 value-added galaxy catalog. 
  The detailed description of the galaxy catalog is in H12.

All the 1529 member galaxies are detected 
  at the \textit{WISE} 3.4 and 4.6 $\micron$
  with signal-to-noise ratio ${\rm (S/N)}\geq3$.
However, there are 1151 (75.3\%) and 552 galaxies (36.1\%) 
  among the members detected at 12 and 22 $\micron$ 
  with ${\rm S/N}\ge3$, respectively.
  We provide physical parameters of the member galaxies used in this paper.
  Table \ref{cat_sdss} lists SDSS-related parameters, 
  and Table \ref{cat_wise} lists \textit{WISE}-related parameters.

\section{Galaxy Groups and Clusters in the A2199 Supercluster}\label{basic}

\begin{deluxetable*}{ccccccccc}
\tablecolumns{9}
\tablewidth{0pc}
\tablecaption{Galaxy Groups/Clusters in the A2199 Supercluster}
\tablehead{
\colhead{} & \colhead{R.A.$_{2000}$} & \colhead{Decl.$_{2000}$} & \colhead{$v$} & \colhead{$\sigma_{p}$} &
   \multicolumn{2}{c}{$\rvir$} & \colhead{$M_{200}$} & \colhead{${\rm log}L_{X}$} \\
\cline{6-7}
\colhead{System} & \colhead{(J2000)} & \colhead{(J2000)} & \colhead{[$\kms$]} & \colhead{[$\kms$]} &
   \colhead{[degree]} & \colhead{[Mpc]} & \colhead{[$10^{14}~M_{\odot}$]} & \colhead{[$h^{-2}{\rm ~erg~s}^{-1}$]} }
\startdata
     A2199 & 16:28:38 & 39:33:05 & 9140$\pm$135 & 675$\pm$35 & 0.73$\pm$0.04 & 1.65$\pm$0.09 & 3.16$\pm$0.48 & 44.1 \\
    A2197W & 16:27:41 & 40:55:40 & 9482$\pm$144 & 537$\pm$22 & 0.58$\pm$0.02 & 1.31$\pm$0.05 & 1.63$\pm$0.19 & 42.5 \\
    A2197E & 16:29:43 & 40:49:12 & 8767$\pm$ 88 & 543$\pm$22 & 0.59$\pm$0.02 & 1.32$\pm$0.05 & 1.68$\pm$0.20 & 42.4 \\
   NRGs385 & 16:17:43 & 34:58:00 & 9321$\pm$100 & 426$\pm$21 & 0.46$\pm$0.02 & 1.04$\pm$0.05 & 0.83$\pm$0.12 & 42.8 \\
   NRGs388 & 16:23:01 & 37:55:21 & 9399$\pm$146 & 549$\pm$33 & 0.60$\pm$0.04 & 1.34$\pm$0.08 & 1.74$\pm$0.30 & 42.3 \\
   NRGs396 & 16:36:50 & 44:13:00 & 9507$\pm$110 & 244$\pm$33 & 0.26$\pm$0.04 & 0.60$\pm$0.08 & 0.17$\pm$0.07 & 42.0 \\
  NGC 6159 & 16:27:25 & 42:40:47 & 9397$\pm$ 94 & 243$\pm$31 & 0.26$\pm$0.03 & 0.59$\pm$0.08 & 0.16$\pm$0.06 & 42.3
\enddata
\label{A2199systems}
\tablecomments{Right ascension and declination are mainly from \citet{rines+01,rines+02}.
In the cases of NRGs385 and the NGC 6159 group, their coordinates are taken from the NASA/IPAC Extragalactic Database.
The uncertainties in $v$ and $\sigma_p$ are determined by the bootstrap re-sampling method.}
\end{deluxetable*}

The A2199 supercluster at $z=0.03$ contains several galaxy groups and clusters.
  \citet{rines+01,rines+02} identified seven galaxy systems including 
  three clusters (A2199, A2197W/E) and four X-ray bright groups 
  (NRGs385, NRGs388, NRGs396, and the NGC 6159 group).
  They derived the physical parameters of these systems.
  Figure \ref{galmap_mir} displays the spatial distribution of three clusters and groups.
  The A2199 supercluster covers a wide range of environment: 
  e.g., $-0.85<{\rm log}\Sigma_5~{\rm (Mpc^{-2})}<2.85$
  if we define the environment by the galaxy surface number density 
  (see \S\ref{local} for details). 
  The density range is wider than the range that we used in the study of A2255 \citep{shim+11}.
  
A2199 is located in the center of the supercluster. 
  It is a typical (rich and regular) cluster of galaxies in the nearby universe.
  The X-ray emission around A2199 is quite symmetric, 
  suggesting that A2199 did not experience recent major merger \citep{rines+01}.
  There is a cooling flow and a radio jet in A2199 
  \citep{markevitch+99,johnstone+02,kawano+03},
  which may be associated with a cD galaxy, NGC 6166 \citep{kelson+02}.
  
Because we have recent redshift data for the supercluster,
  we recalculate some physical parameters of these systems.
For example, we compute the mean redshifts, 
  $v~[\kms]$ (biweight location of \citealt{beers90}),
  using the galaxies at $R\leq0.25~h^{-1}$ Mpc of each system.
The radius for computing the mean redshifts is small enough 
  so that the galaxy systems do not overlap each other,
  as used in \citet{rines+02}.
We also compute the projected velocity dispersions, 
  $\sigma_p~[\kms]$ (biweight scale of \citealt{beers90}), 
  using the galaxies at $R\leq1.0~h^{-1}$ Mpc.
We use this radius, smaller than the one used in \citet{rines+02} 
  (i.e., 1.5 $h^{-1}$ Mpc),
  to avoid the overlap between the systems.
If we use $R=1.5~h^{-1}$ Mpc aperture, 
  $\sigma_p$ changes within the uncertainty.

Our estimates of $v$ for the systems in the A2199 supercluster
  are in the range $8767-9507~\kms$;
  these agree well with previous measurements in \citet{rines+02}.
On the other hand, our $\sigma_p$ ranges from $244~\kms$ to $675~\kms$;
  these are on average $\sim20\%$ smaller than previous measurements 
  in \citet{rines+02}.
This difference seems to result mainly from the difference in
  the data.
For example, $\sigma_p$ of A2199 in this study is $675\pm35~\kms$.
This is smaller than $796^{+38}_{-33}~\kms$ in \citet{rines+02}, 
  but is similar to $676^{+37}_{-32}~\kms$ 
  in \citet{rines+06} who used the same SDSS data.
Table \ref{A2199systems} lists our measurements 
  of $v$ and $\sigma_p$ for each system.

To estimate the size of each system, 
  we convert $\sigma_p$ into $\rvir$ 
  (approximately the virial radius) 
  using the formula given in \citet{carlberg97}.
We then determine the virial mass ($M_{200}$) of each system 
  using the $M_{200}-\sigma_p$ relation given in \citet{rines+13}:
  $M_{200}~[10^{14} M_{\odot}]=0.093\times(\sigma_p/200)^{2.90\pm0.15}$.
The derived values of $\rvir$ and $M_{200}$ of A2199 are $1.65\pm0.09$ Mpc and
  $(3.16\pm0.48)\times10^{14} M_{\odot}$, respectively.
These values agree well with the estimates in \citet{rines+06}:
  $\rvir=1.44$ Mpc and $M_{200}=(3.41\pm1.10$)$\times10^{14}~M_{\odot}$.
The $M_{200}$ for other systems in the A2199 supercluster 
  are in the range $0.15-2\times10^{14}~M_{\odot}$,
  similar to the masses of galaxy groups at $z<0.1$ \citep{rines+10}.
Table \ref{A2199systems} lists $\rvir$ and $M_{200}$ 
  of the groups and clusters in the A2199 supercluster.

\section{Integrated SFRs of Galaxy Systems in the A2199 Supercluster}\label{isfr}

\begin{deluxetable*}{ccccccccccc}
\tablecolumns{11}
\tablewidth{0pc}
\tablecaption{$\Sigma{\rm SFR}$ and $\Sigma{\rm SFR}/M_{200}$ of Groups/Clusters in A2199 Supercluster}
\tablehead{
\colhead{} & \multicolumn{6}{c}{$R\leq 0.5\rvir$} & & \multicolumn{3}{c}{$R\leq \rvir$} \\
\cline{2-7} \cline{9-11}
\colhead{Galaxy}
& \colhead{} & \colhead{$\Sigma {\rm SFR}~(>2.0)$} & \colhead{$\Sigma {\rm SFR} / M_{200}$} & 
  \colhead{} & \colhead{$\Sigma {\rm SFR}~(>0.5)$} & \colhead{$\Sigma {\rm SFR} / M_{200}$} &
& \colhead{} & \colhead{$\Sigma {\rm SFR}~(>0.5)$} & \colhead{$\Sigma {\rm SFR} / M_{200}$} \\
\colhead{System}
& \colhead{$N_{\rm gal}$} & \colhead{$[M_{\odot}~{\rm yr}^{-1}]$} & \colhead{$[10^{-14}~{\rm yr}^{-1}]$} & 
  \colhead{$N_{\rm gal}$} & \colhead{$[M_{\odot}~{\rm yr}^{-1}]$} & \colhead{$[10^{-14}~{\rm yr}^{-1}]$} &
& \colhead{$N_{\rm gal}$} & \colhead{$[M_{\odot}~{\rm yr}^{-1}]$} & \colhead{$[10^{-14}~{\rm yr}^{-1}]$}}
\startdata
    A2199 &  3 & $15.09\pm 3.98$ & $4.78\pm 1.45$ & 15 & $24.67\pm 4.16$ & $7.81\pm 1.76$ & & 34 & $51.97\pm 5.56$ & $16.44\pm 3.04$ \\
   A2197W &  0 & ...             & ...             & 12 & $10.68\pm 1.41$ & $6.56\pm 1.17$ & & 33 & $40.01\pm 3.44$ & $24.57\pm 3.61$ \\
   A2197E &  1 & $3.32\pm 1.36$ & $1.97\pm 0.84$ & 12 & $13.37\pm 1.90$ & $7.95\pm 1.47$ & & 37 & $48.82\pm 4.24$ & $29.03\pm 4.24$ \\
  NRGs385 &  0 & ...             & ...             &  6 & $3.73\pm 0.63$ & $4.49\pm 1.00$ & & 12 & $12.50\pm 1.77$ & $15.03\pm 3.02$ \\
  NRGs388 &  0 & ...             & ...             &  4 & $5.87\pm 1.25$ & $3.38\pm 0.93$ & & 15 & $16.66\pm 2.05$ & $9.60\pm 2.05$ \\
  NRGs396 &  0 & ...             & ...             &  2 & $1.44\pm 0.42$ & $8.74\pm 4.28$ & &  6 & $8.90\pm 2.22$ & $53.83\pm25.11$ \\
 NGC 6159 &  0 & ...             & ...             &  2 & $1.20\pm 0.35$ & $7.33\pm 3.46$ & &  2 & $1.20\pm 0.35$ & $7.33\pm 3.46$
\enddata
\label{table_ssfr}
\end{deluxetable*}

\begin{figure*}
\centering
\includegraphics[scale=0.6]{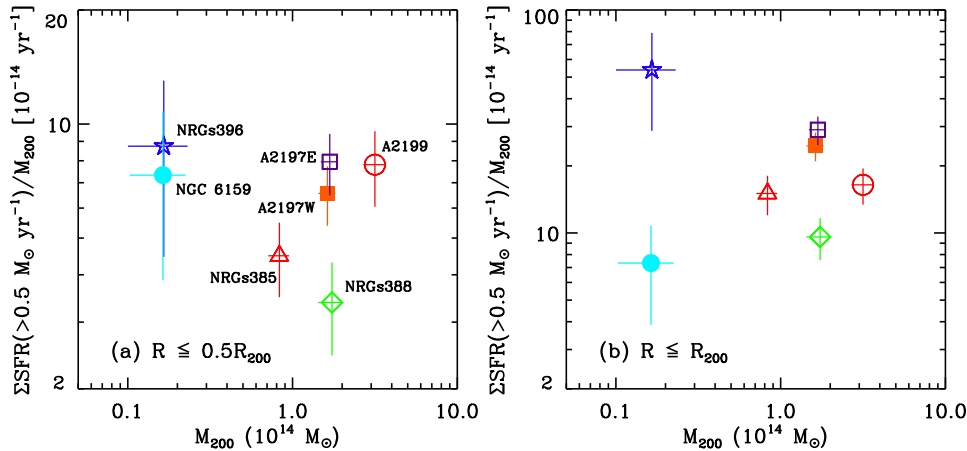}
\caption{$\Sigma{\rm SFR}/M_{200}$ for (a) $R\le0.5\rvir$ and (b) $R\leq\rvir$ of 
  galaxy groups/clusters as a function of $M_{200}$
  using galaxies with ${\rm SFR_{WISE}}>0.5~M_{\odot}~{\rm yr}^{-1}$.}
\label{m200_ssfr}
\end{figure*}

To study the dependence of the star formation activity of clusters on cluster mass,
  many studies have used the mass-normalized integrated SFRs of clusters 
  (i.e., $\Sigma{\rm SFR}/M_{200}$)
  as a proxy for the global SFRs
  \citep[e.g.,][]{finn+05,geach+06,bai+07,bai+09,chung+10,
  koyama+10,shim+11,biviano+11,chung+11}.
In this section, we study this dependence for the galaxy systems 
  in the A2199 supercluster
  using the \textit{WISE} data.
  
To compute the integrated SFR of each system,
  we use the SFRs of galaxies 
  converted from total IR ($8-1000~\micron$) luminosities ($L_{\rm IR}$).
We compute the total IR luminosities of galaxies 
  using the \emph{WISE} 22 $\micron$ flux densities
  and the spectral energy distribution templates of \citet{chary+01}.
Following H12, we use the relation in \citet{kennicutt98} 
  for the conversion of the total IR luminosities into SFRs 
  with the assumption of a \citet{salpeter55} initial mass function
  with a power law index $x=2.35$:
  ${\rm SFR_{WISE}}~(M_{\odot}~{\rm yr^{-1}})=1.72\times10^{-10}~L_{\rm IR} (L_{\odot})$.
These IR-based SFRs (${\rm SFR_{WISE}}$) of galaxies agree well with
  those derived from the optical spectra 
  (H12; \citealt{hwang+12shels,lee+13}).

To have a fair comparison with other studies 
  \citep[e.g.,][]{finn+05,bai+07,bai+09},
  we first use the same selection criteria adopted in other studies 
  to compute the integrated SFRs 
  (i.e., ${\rm SFR_{WISE}}>2~M_\odot~{\rm yr}^{-1}$
  and $R\leq0.5\rvir$).
We do not include active galactic nucleus (AGN)-host galaxies 
  such as Seyferts, LINERs, 
  Type I AGNs with broad Balmer lines (FWHM $> 1000~\kms$) 
  and MIR-selected AGNs ($[3.4]-[4.6]>0.44$, H12)
  because their $L_{\rm IR}$ (based on the \emph{WISE} 22 $\micron$ flux densities)
  can be dominated by AGN rather than 
  by star formation \citep{brand+09,lee12akari}.
However, we include composite galaxies
  because the AGN contribution to the 22 $\micron$ flux densities 
  in these galaxies is small ($\sim6\%$) \citep{lee12}.

\begin{figure*}
\centering
\includegraphics[scale=0.6]{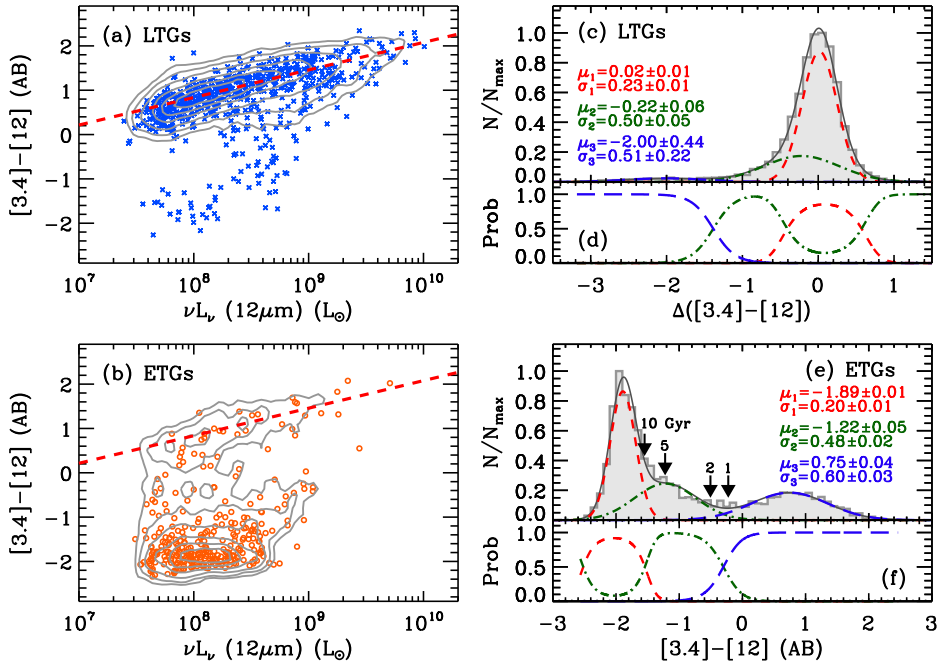}
\caption{(a) MIR color-luminosity diagram for (a) late-type and (b) early-type galaxies.
Contours represent number density distribution of galaxies at $0.025<z<0.035$, while
color symbols are member galaxies in the A2199 supercluster. Dashed lines are 
the ``MIR SF sequence'' that is a linear fit for late-type galaxies with 
SF nuclei, which is defined in \citet{hwang+12}.
(c) Histogram of $\Delta([3.4]-[12])$, the deviation of $[3.4]-[12]$ from 
the MIR SF sequence fit, for late-type galaxies.
We use the Gaussian mixture modeling \citep[GMM,][]{muratov+10} to 
decompose the histogram with multi-Gaussian functions. We overplot 
three Gaussians and their sum, and list the mean and standard deviation 
of each Gaussian. 
(d) the probability distribution from the GMM that indicates which Gaussian 
galaxies should belong to. 
(e) Histogram of $[3.4]-[12]$ for early-type galaxies, and three Gaussians 
and the sum obtained from the GMM.
Arrows represent the predicted $[3.4]-[12]$ colors of early-type galaxies 
with different mean stellar ages from the SSP models \citep{piovan+03}. 
(f) The Probability of $[3.4]-[12]$ belonging to each Gaussian.}
\label{lumw3w13}
\end{figure*}

As listed in Table \ref{table_ssfr}, 
  we could obtain $\Sigma{\rm SFR}/M_{200}$ only for 
  two galaxy systems: 
  $4.78\pm1.45$ for A2199 and $1.97\pm0.84$ for A2197E in unit of 
  $10^{-14}~{\rm yr}^{-1}$. 
These values are similar to other clusters at similar redshifts
  \citep[e.g., Coma and A1367,][]{bai+07}.
However, we could not determine $\Sigma{\rm SFR}/M_{200}$
  for the other systems because there are no galaxies 
  satisfying the selection criteria
  (i.e., ${\rm SFR_{WISE}}>2~M_\odot~{\rm yr}^{-1}$
  and $R\leq0.5\rvir$). 
It also should be noted that $\Sigma{\rm SFR}/M_{200}$
  for A2199 and A2197E
  are based only on a small number of galaxies
  (i.e., three and one galaxies for A2199 and A2197E, respectively).
    
Thus, we change the selection criteria to increase the sample size:
  ${\rm SFR_{WISE}}>0.5~M_\odot~{\rm yr}^{-1}$
  and $R\leq0.5\rvir$.
Most galaxies with ${\rm SFR_{WISE}}>0.5~M_\odot~{\rm yr}^{-1}$ are
  detected at 22 $\micron$ with ${\rm S/N}_{22\micron}\geq3$ (94\%),
  suggesting no significant bias due to incompleteness.
This limit is much lower than the one used in a recent study
  based on similar \textit{WISE} data 
  (e.g., $4.6~M_\odot~{\rm yr}^{-1}$ in \citealt{chung+11}).
  
Table \ref{table_ssfr} lists 
  $\Sigma{\rm SFR}$ and $\Sigma{\rm SFR}/M_{200}$ of each system
  based on samples of galaxies
  with ${\rm SFR_{WISE}}>0.5~M_\odot~{\rm yr}^{-1}$.
We derive the values 
  for two cases: $R\leq0.5\rvir$ and $R\leq\rvir$.
Figure \ref{m200_ssfr}(a) shows that 
  $\Sigma{\rm SFR}/M_{200}$ for $R\leq0.5\rvir$
  does not depend on $M_{200}$. 
Similarly, $\Sigma{\rm SFR}/M_{200}$ for $R\leq\rvir$ in panel (b)
  does not show $M_{200}$ dependence. 
The Spearman's coefficient for the data in panel (a) is $-0.18$, while 
  that in panel(b) is 0.07. 
  In addition, the probability of obtaining the correlation by chance 
  is over 70\% in the both case.
  These results suggest that 
   there is no significant correlation between 
   $\Sigma{\rm SFR}/M_{200}$ and $M_{200}$.

\section{MIR Properties of Galaxies in Galaxy Systems of the A2199 Supercluster}\label{mirc}

To compare the star formation activity of galaxy systems in the A2199 supercluster in detail,
  we investigate the MIR properties of the galaxies 
  in each system in this section.
We first divide galaxies into three MIR classes 
  in the MIR color-luminosity diagram (\S \ref{class}), 
  and then compute the fraction of galaxies in each MIR class
  for the galaxy groups/clusters in the A2199 supercluster 
  (\S \ref{mirp}). 
We also examine the environmental dependence
  of the MIR properties of galaxies in \S \ref{local}.

\subsection{Galaxy Classification in the MIR color-luminosity Diagram}\label{class}

The MIR colors such as \emph{AKARI} $N3-S11$ or
  \emph{WISE} $[3.4]-[22]$, $[4.6]-[12]$
  are useful indicators of the specific SFRs and
  of the presence of intermediate-age
  stellar populations 
  (e.g., \citealt{lee+09,ko+09,ko+12,ko+13,shim+11,donoso+12},H12).
\emph{WISE} $[3.4]-[12]$ colors are also
  a good indicator of the specific SFRs
  because 3.4 and 12 $\micron$ luminosities, respectively, 
  trace stellar masses and SFRs \citep{donoso+12,hwang+12shels}.
The other \emph{WISE} colors including 
  $[4.6]-[12]$, $[3.4]-[22]$, and $[4.6]-[22]$
  can also be used.
However, we use $[3.4]-[12]$ in this study
  because 3.4 $\micron$ probes the old stellar components better
  than than 4.6 $\micron$, and
  12 $\micron$ sensitivity is better than 22 $\micron$.

The left panels in Figure \ref{lumw3w13} show
  the distribution of galaxies in the $[3.4]-[12]$ versus 
  12 $\micron$ luminosity diagram. 
As shown in H12, 
  late-type SF galaxies form a linear ``MIR SF sequence'', while 
  most early-type galaxies form a ``MIR blue cloud'' 
  with low 12 $\micron$ luminosities.
For robust classification of the galaxies in this diagram,
  we use a large sample of SDSS galaxies at $0.025<z<0.035$ 
  including the A2199 supercluster galaxies.
The contours in Figure \ref{lumw3w13}(a) and (b)
  indicate the number density distributions of 
  early- and late-type galaxies, respectively.
The linear fit to this large galaxy sample gives a relation, 
  \begin{equation}
  [3.4]-[12]={\rm log}(\nu L_{\nu}(12\micron))\times(0.62\pm0.01)-(4.13\pm0.05).
  \end{equation}
This relation is similar to the one derived in H12, 
  but based on a much larger sample.

We plot the distribution of vertical offsets
  of late-type galaxies
  from the MIR SF sequence, $\Delta([3.4]-[12])$, in panel (c).
The histogram is negatively skewed.  
We use the Gaussian mixture modeling \citep[GMM,][]{muratov+10}
  to decompose the histogram with multi-Gaussian functions.
The histogram is well described by the sum of three Gaussians
  rather than the sum of two Gaussians.
We overplot the three Gaussians with their sum,
  and list the mean and standard deviation of each Gaussian. 

Figure \ref{lumw3w13}(d) shows the probability distribution
  from the GMM that indicates
  which Gaussian galaxies should belong to.
At $\Delta([3.4]-[12])<-1.38$,
  the probability that galaxies belong to the left Gaussian 
  (long-dashed line) is larger than 50\%. 
At $-1.38<\Delta([3.4]-[12])<-0.41$,
  the probability that galaxies belong to the middle Gaussian 
  is over 50\%.
We adopt $\Delta([3.4]-[12])=-1.38$  
  to separate the galaxies in the left tail
  from the majority of late-type, SF galaxies.
  
On the other hand, the majority of early-type galaxies
  are distributed around $[3.4]-[12]=-2$, 
  which is the MIR blue cloud. 
Some early-type galaxies have MIR colors,
  much redder than the MIR blue cloud.
Figure \ref{lumw3w13}(e) shows the distribution
  of $[3.4]-[12]$ colors for early-type galaxies at $0.025<z<0.035$. 
The histogram is well described by the sum of three Gaussians.
The Gaussian in the left covers a narrow color range 
  centered at $[3.4]-[12]\simeq-1.9$ (i.e., MIR blue cloud). 
The Gaussians in the middle and right
  are for those in the green valley (to be defined in this section)
  and in the MIR SF sequence, respectively.

\begin{figure}
\centering
\includegraphics[scale=0.65]{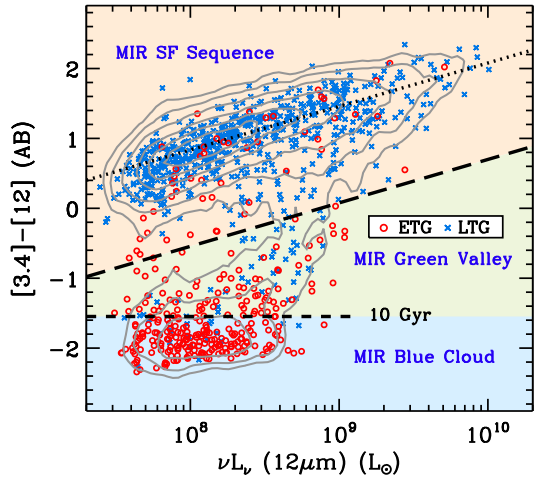}
\caption{MIR color-luminosity diagram for galaxies with ${\rm S/N_{12\micron}}\ge3$.
  Contours represent the number density distribution of galaxies at $0.025<z<0.035$, 
  while circles and crosses are early- and late-type member galaxies of 
  the A2199 supercluster, respectively. 
  We divide galaxies into three classes: MIR SF sequence galaxies,
  MIR green valley galaxies, and MIR blue cloud galaxies. 
  Long and short dashed lines are demarcation lines for this MIR galaxy classification.
  The latter corresponds to the the color of 10 Gyr stellar populations \citep{piovan+03}.  
  A dotted line indicates a linear fit for the MIR SF sequence.}
\label{mirclass}
\end{figure}
  
We overplot the predicted $[3.4]-[12]$ colors  
  from the single stellar population models (SSP) of \citet{piovan+03}.
This model considers the MIR emission from dusty circumstellar envelops 
  of asymptotic giant branch stars when constructing
  spectral energy distribution of early-type galaxies.
Four arrows indicate
  mean stellar ages of 1, 2, 5, and 10 Gyr
  ($[3.4]-[12]=-0.22$, $-0.51$, $-1.22$, and $-1.55$).
This suggests that the Gaussian in the left 
  mainly traces the galaxies with stellar populations
  older than 10 Gyrs,
  while the Gaussian in the middle traces the galaxies with
  intermediate-age stellar populations.
  
To separate the early-type galaxies with intermediate and old
  stellar populations,
  we adopt a cut of $[3.4]-[12]=-1.55$ that 
  corresponds to the color of 10 Gyr stellar populations. 
This color cut is the one where the probability 
  that the galaxies belong to the Gaussian in the left
  becomes larger than 50\%, as shown in panel (f). 

Based on the results in Figure \ref{lumw3w13}, 
 we divide the galaxies into three MIR classes 
 in the MIR color-luminosity diagram as shown in Figure \ref{mirclass}. 
We use two demarcation lines, 
  $\Delta([3.4]-[12])=-1.38$ from the MIR SF sequence
 and $[3.4]-[12]=-1.55$.
We call the galaxies at $\Delta([3.4]-[12])>-1.38$ from the MIR SF sequence
 ``MIR SF sequence galaxies'', 
 and those with $[3.4]-[12]<-1.55$ ``MIR blue cloud galaxies''. 
Then the galaxies between the two lines are ``MIR green valley galaxies''. 
 
Among 997 member galaxies in this diagram, 
  the number of MIR SF sequence, MIR green valley, and MIR blue cloud galaxies 
  are 678 ($68.0\%$), 126 ($12.6\%$), and 193 ($19.4\%$), respectively.
The majority (93\%, 628/678) of MIR SF sequence galaxies 
  are morphologically late types.
In contrast, MIR blue cloud galaxies are 
  mainly (90\%, 174/193) early types. 
Two-thirds of MIR green valley galaxies are early-type galaxies (68\%, 86/126),
  and one third of them are late types (40/126). 

\begin{figure}
\centering
\includegraphics[scale=0.65]{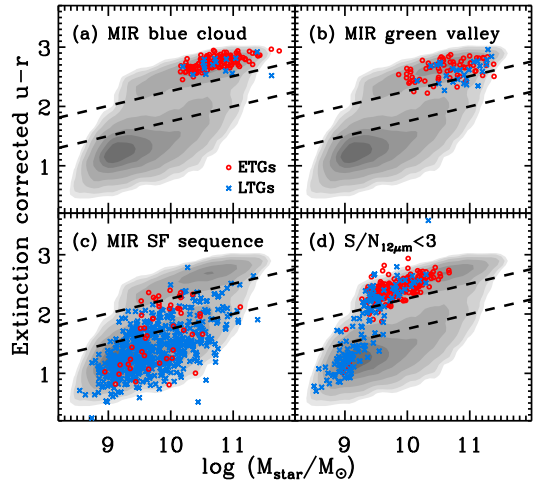}
\caption{The extinction-corrected $(u-r)$ color vs. stellar mass diagram.
Gray contours represent the number density distribution of galaxies 
at $0.025<z<0.035$. Circles and crosses represent early- and late-type 
A2199 supercluster members segregated by their MIR classes:
(a) MIR blue cloud galaxies, (b) MIR green valley galaxies, (c) MIR SF
sequence galaxies, and (d) 12 $\micron$ undetected galaxies (${\rm S/N_{12\micron}}<3$).
Dashed lines indicate the optical green valley defined in \citet{schawinski+14}.}
\label{opt_CMD}
\end{figure}

\begin{figure*}
\centering
\includegraphics[scale=0.6]{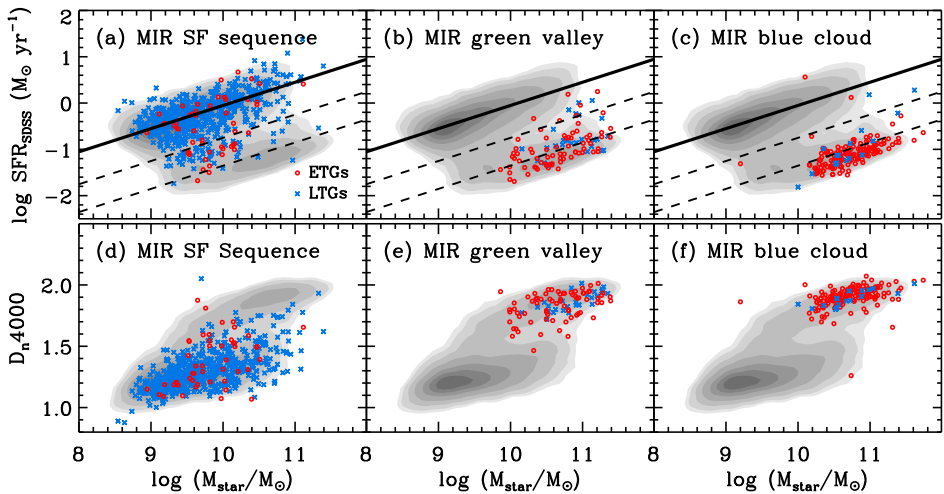}
\caption{Top: stellar mass vs. $\sfropt$ diagram. Contours represent number density distribution
of galaxies at $0.025<z<0.035$. Circles and crosses represent early- and late-type member galaxies
of the A2199 supercluster segregated by their MIR classes: (a) MIR SF sequence galaxies,
(b) MIR green valley galaxies, and (c) MIR blue cloud galaxies.
Solid lines are the linear fit to galaxies with SF nuclei, while 
dashed lines indicate $1/5$ and $1/20$ levels, respectively.
Bottom: stellar mass vs. $D_n4000$ distribution of (d) MIR SF sequence, (e) MIR green valley, 
and (f) MIR blue cloud galaxies.}
\label{stmass_sfr}
\end{figure*}
  
To compare the MIR classification with that in the optical bands,
  we plot the galaxies in each MIR class 
  in the optical color-mass diagram. 
Figure \ref{opt_CMD} shows an extinction-corrected $u-r$ color 
  as a function of stellar mass. 
To correct for dust extinction, 
  we use E(B-V) values from the stellar continuum fits of \citet{oh+11}
  and apply the reddening law of \citet{cardelli+89}.
Grayscale contours indicate the number density of 
  the SDSS galaxies at $0.025<z<0.035$. 
Two dashed lines define the optical green valley 
  adopted from \citet{schawinski+14}.
  
As expected, MIR SF sequence galaxies are mainly distributed 
  in the optical blue cloud (see panel c).
Some of them are in the optical green valley.
MIR blue cloud galaxies are distributed in the optical red sequence (panel a).
Interestingly, the majority of MIR green valley galaxies 
  roughly follow the optical red sequence despite 
  a scatter in color (panel b).
These are consistent with the results in other studies \citep{ko+12, walker+13}.
  
The stellar mass range of MIR green valley galaxies is nearly the same 
  as that of MIR blue cloud galaxies 
  (${\rm log}~(M_{\rm star}/M_{\odot})>10$).
This means that the discrimination between MIR blue cloud and 
  MIR green valley galaxies based only on optical parameters is difficult.
It is interesting to note that 
  the MIR green valley galaxies are not the same as 
  the optical green valley galaxies.
  The optical green valley galaxies are included in the MIR SF sequence class. 
Panels (a-c) also show that 
  the distributions of early- and late-type galaxies in each panel
  are not significantly different.
    
In panel (d) we plot the distribution of 12 $\micron$ 
  undetected galaxies (i.e., ${\rm S/N}_{12\micron}<3$);
  they are not included in the MIR classification.
They are distributed either in the optical blue cloud or in the red sequence. 
They appear to have stellar masses lower than other MIR classes
  in a fixed morphology.
The majority ($\sim80\%$) of 
  the 12 $\micron$ undetected galaxies have no strong emission lines 
  in their optical spectra; most of their spectral types are undetermined.
These results suggest that 
  they are a mixture of young galaxies with small stellar masses
  and of relatively massive galaxies that are quiescent; 
  both of them have 12 $\micron$ flux densities,
  not high enough to be detected with \textit{WISE}. 

We also show the galaxies in each MIR class
  in the plot of SFR derived in the optical band and stellar mass
  (see top panels of Figure \ref{stmass_sfr}).
We use $\sfropt$ adopted from the MPA/JHU DR7 
  value-added galaxy catalog \citep{brinchmann+04}.
The grayscale number density contours
  based on the galaxies at $0.025<z<0.035$
  show two distinctive sequences.
The upper sequence (solid lines) is for SF galaxies, 
  which is called a ``main sequence'' of SF galaxies
  \citep{noeske+07,elbaz+11}.
The lower sequence is for those with SFRs much smaller than
  the main sequence galaxies.
  
As expected, MIR SF sequence galaxies are well 
  distributed along the main sequence (panel a). 
In contrast, MIR blue cloud galaxies are 
  on the sequence in the bottom.
Similarly, MIR green valley galaxies 
  are mainly found on the sequence in the bottom.
The SFRs of MIR green valley galaxies 
  appear to be slightly higher than 
  those for MIR blue cloud galaxies in a give stellar mass.
However, the difference is not significant. 
  
We also examine the distribution of $D_n4000$ 
  for the three MIR classes in bottom panels. 
The $D_n4000$ is a useful measure of mean stellar age of galaxies
  \citep{balogh99,shim+11}.
Galaxies with $D_n4000<1.5$ contain young stellar populations
  with $\lesssim1$ Gyr \citep{kauffmann+03}.
On the other hand, $D_n4000$ larger than 1.5 indicates mean stellar age
  larger than 1 Gyrs \citep{kranz+10}.
As expected, most MIR SF sequence galaxies have $D_n4000$ smaller than 1.5. 
In contrast, $D_n4000$ of
  the majority of the MIR green valley and MIR blue cloud galaxies 
  is larger than 1.5.
The mean $D_n4000$ of the MIR green valley galaxies ($1.84\pm0.03$) is not different 
  significantly from that of the MIR blue cloud galaxies ($1.90\pm0.01$).
  However, the rms of $D_n4000$ for the MIR green valley galaxies is larger ($0.10\pm0.01$)
  than that for the MIR blue cloud galaxies ($0.07\pm0.01$), 
  suggesting that the dispersion of mean stellar ages for the MIR green valley galaxies 
  is larger than that for the MIR blue cloud galaxies. 
  However, the difference in mean stellar ages between the two populations 
  is more evident in the MIR color space.

\begin{figure*}
\centering
\includegraphics[scale=0.6]{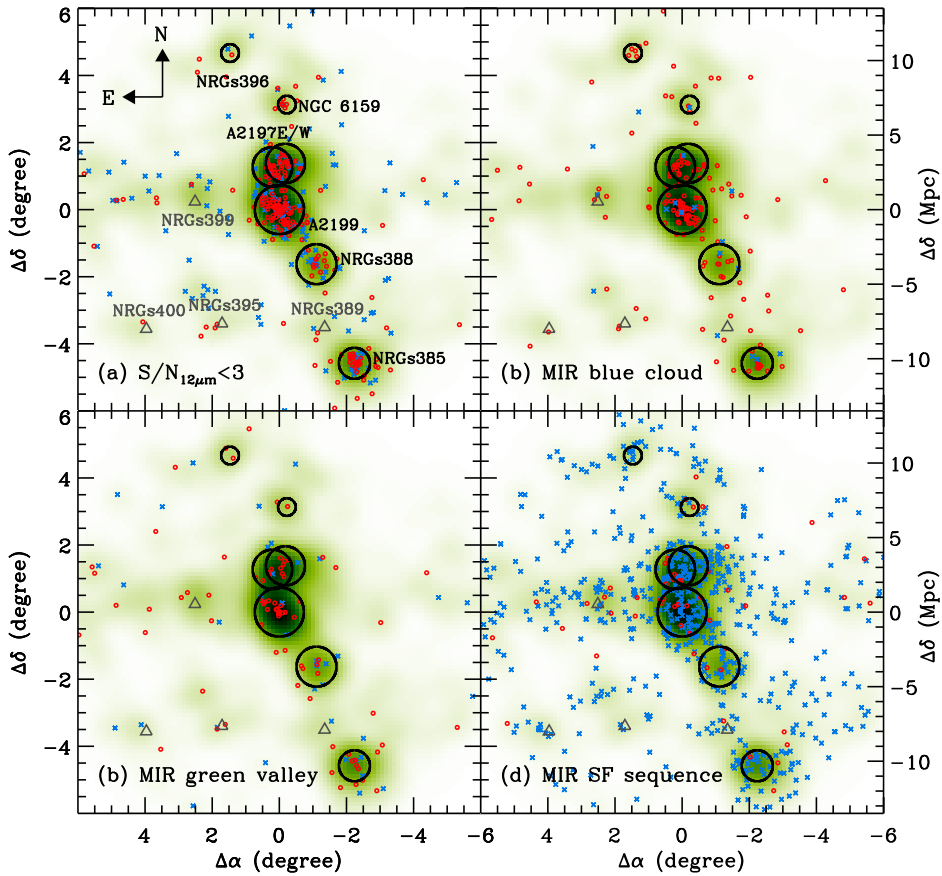}
\caption{Spatial distribution of (a) the member galaxies undetected at 12$\micron$ (${\rm S/N}_{12\micron}<3$),
  (b) MIR blue cloud galaxies, (c) MIR green valley galaxies, and
  (d) MIR SF sequence galaxies in the A2199 supercluster.
  The green and white represent the galaxy number density distribution.
  Large circles represent the positions of X-ray bright clusters and groups, 
  and their sizes are proportional to virial radii ($\rvir$).
  Triangles represent the positions of four X-ray faint groups. 
  Small circles and crosses represent early- and late-type galaxies, respectively.}
\label{galmap_mir}
\end{figure*}

\subsection{Galaxies in groups/clusters and their MIR Classes}\label{mirp}

In this section 
  we examine the spatial distribution of three MIR classes
  in the A2199 supercluster.
We compute the fraction of 
  each MIR class in galaxy groups/clusters, 
  and examine the variation of the fraction with the mass of each system.

We first show the spatial distribution of galaxies in each MIR class 
  over the entire region of the A2199 supercluster 
  in Figure \ref{galmap_mir}.
We mark the positions of three clusters (A2199, A2197W/E) and four X-ray 
  bright groups (NRGs385, NRGs388, NRGs396, and the NGC 6159 group) as thick circles.
The size of circles is proportional to their virial radii ($\rvir$). 
We also mark the positions of four X-ray faint groups 
  \citep[NRGs389, NRGs395, NRGs399, and NRGs400,][]{rines+01} as triangles.
To better show the clustering of galaxies, 
  we overplot the grayscale galaxy number density map using all the member
  galaxies regardless of their MIR classes in each panel.
  
The 12 $\micron$ undetected galaxies (panel a) are mainly found 
  in high-density regions. 
Early-type galaxies are highly concentrated in central regions 
  of groups/clusters.
Note that these early-type galaxies are in the 
  optical red sequence with low stellar masses. 
Similarly, MIR blue cloud galaxies (panel b) and 
  MIR green valley galaxies (panel c) 
  are mainly found in groups and clusters. 
In contrast, MIR SF sequence galaxies are distributed 
  over the entire supercluster region. 
Some of them are in groups and clusters, 
  but they do not show a central concentration 
  like MIR blue cloud and MIR green valley galaxies.
Four X-ray faint groups do not show any significant clustering 
  of galaxies. 
  
We then calculate the fraction of each MIR class 
  in the X-ray bright groups and clusters.  
We use the galaxies in the inner region ($R\leq0.5\rvir$) and 
  in the outer region ($0.5\rvir<R\leq\rvir$) separately.
For comparison, we also calculate the fraction of each MIR class  
  for the galaxies not associated with any 
  group or cluster (i.e., $R>3\times\rvir$) in the supercluster region;
  these are the galaxies in ``underdense regions''.  
We use only 12 $\micron$ detected galaxies for computing fractions. 
If we use all the galaxies to compute the fractions by 
  including the 12 $\micron$ undetected galaxies, 
  the results do not change much. 

\begin{figure*}
\centering
\includegraphics[scale=0.6]{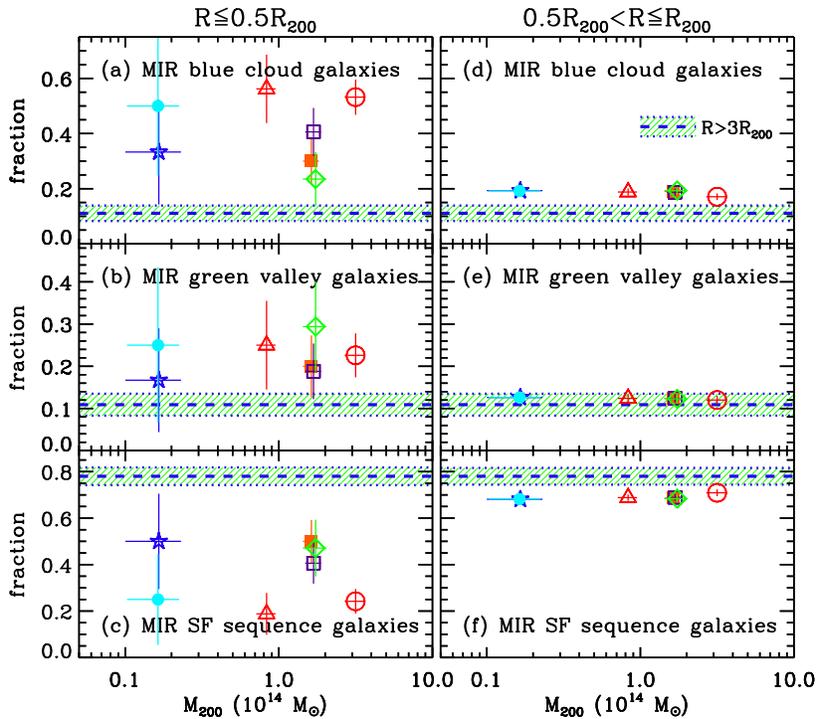}
\caption{The fraction of (top) MIR blue sequence, (middle) MIR green valley, 
  and (bottom) MIR SF sequence galaxies in the groups/clusters as a function of 
  $M_{200}$. We use galaxies in the inner region ($R\leq0.5R_{200}$, left) and 
  the outer region ($0.5R_{200}<R\leq R_{200}$, right), separately, of groups/clusters.
  The symbols are the same as Figure \ref{m200_ssfr}. 
  Hatched regions represent the fraction of each MIR class in the underdense region.}
\label{m200_mirfrac}
\end{figure*}
  
The left panels of Figure \ref{m200_mirfrac} 
  show the fraction of each MIR class in the inner region 
  as a function of $M_{200}$ of each galaxy system.
The fractions of MIR blue cloud galaxies (panel a) and 
  of MIR SF sequence galaxies (panel c) 
  vary from $\sim20\%$ to $\sim60\%$, 
  but do not show any dependence on $M_{200}$.
The fractions of MIR green valley galaxies in the galaxy systems
  are similar ($\sim25\%$). 

Similarly, the right panels for the galaxies in the outer region of 
  groups/clusters show no dependence of the fraction on $M_{200}$.
One unique point is that the fraction of each MIR class 
  is nearly the same for all groups and clusters.
Although we calculate the fraction using the galaxies 
  at $R\leq\rvir$ of each group and cluster
  (i.e., sum of the left and right panels),
  the $M_{200}$ dependence of the fraction is not apparent. 

On the other hand, panel (a) shows that 
  the fractions of MIR blue cloud galaxies in the inner region of groups/clusters
  are significantly higher than in the underdense region (11.1\%).
However, the fractions in the outer region (panel d) are 
  on average $\sim$18.5\%, close to the fraction in the underdense region. 
The fraction of MIR green valley galaxies in the underdense region is 10.9\% 
This value is smaller than those in the inner region (panel b), but 
  is comparable to those in the outer region (panel e). 
The fractions of MIR SF sequence galaxies
  in the inner region (panel c) 
  are much smaller than that in the underdense region (78\%). 
However, the fractions in the outer region are close to 
  the value in the underdense region (panel f).

Thus Figure \ref{m200_mirfrac} 
  suggests that massive galaxies with low SFRs 
  (i.e., MIR blue cloud galaxies and MIR green valley galaxies)
  are mainly in the inner regions of groups/clusters
  regardless of their $M_{200}$. 
Some of them are found
  even in the low density regions (i.e., the underdense region), 
  but their fraction is not significant (up to $\sim20\%$).

\begin{figure}
\centering
\includegraphics[scale=0.65]{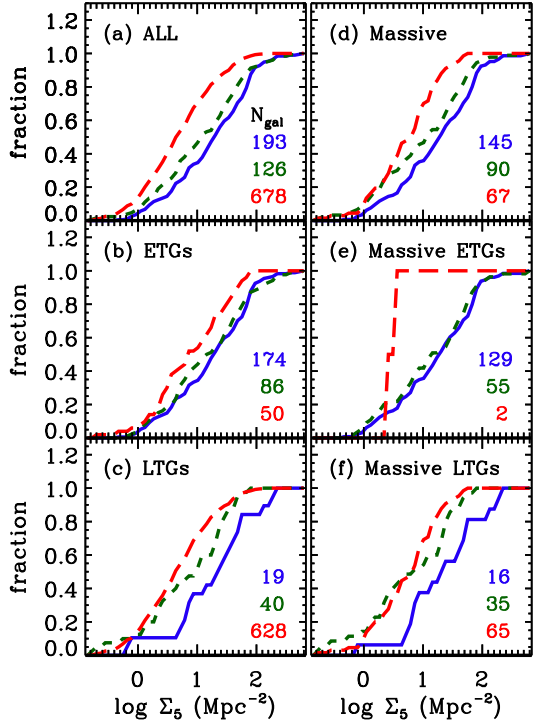}
\caption{Cumulative distribution function of the three MIR classes as a function of 
  the surface galaxy number density 
  ($\Sigma_5$) for MIR blue cloud galaxies (solid line), MIR green valley galaxies 
  (short dashed line), and MIR SF sequence galaxies (long dashed line). 
   The galaxy sample used in each panel is (a) all 12 $\micron$ detected galaxies, 
  (b) early-type galaxies, (c) late-type galaxies, 
  (d) massive galaxies greater than ${\rm log}~(M_{\rm star}/M_{\odot})>10.5$,
  (e) massive early-type galaxies, and (f) massive late-type galaxies. 
  We list the number of three MIR class galaxies used in each panel.}
\label{cum_sig5}
\end{figure}

\subsection{Environmental Dependence based on $\Sigma_5$ and $R/\rvir$}\label{local}

To further examine the environmental dependence of
  the MIR classes in the A2199 supercluster, 
  we use two environmental indicators in this section; 
  the surface number density ($\Sigma_5$) and 
  the cluster-(or group-)centric distance normalized by virial 
  radius ($R/\rvir$).
The surface number density, $\Sigma_5$, 
  is defined by $\Sigma_5=5(\pi D^2_{p,5})^{-1}$,
  where $D_{p,5}$ is the projected distance to the 5th-nearest neighbor.
The fifth-nearest neighbor to each galaxy is identified
  in the sample of A2199 supercluster member galaxies.
We compare the cumulative distribution of each MIR class
  as function of two environmental indicators. 
We also use the Kolmogorov-Smirnov two-sample test 
  to quantify the difference in the distribution if any.
 
Figure \ref{cum_sig5} shows the cumulative distribution
  of each MIR class as a function of $\Sigma_5$. 
Panel (a) shows the case for all the 12 $\micron$ detected galaxies. 
MIR SF sequence galaxies (red long dashed line) 
  tend to be in low-density regions, 
  and MIR blue cloud galaxies (blue solid line) prefer high-density regions.
MIR green valley galaxies (green dashed line) are between the two.
This result is consistent with the expectations from 
  the morphology- and/or SFR-density relation 
  \citep[e.g.,][]{dressler80,lewis+02,park+09,hwang+10}. 

Panel (b) shows the same $\Sigma_5$ distribution of each MIR class,
  but only for early-type galaxies.
It shows that the difference between MIR blue cloud and 
  MIR green valley galaxies is noticeably reduced.
The confidence level to reject the hypothesis that the two distributions
  are extracted from the same parent population
  changes from 92\% in panel (a) into 52\% in panel (b).
This suggests that $\Sigma_5$ of MIR green valley 
  early-type galaxies do not differ significantly 
  from that of MIR blue cloud early-type galaxies.
On the other hand, the cumulative distribution of MIR SF sequence 
  early-type galaxies is still distinguishable from 
  those of MIR blue cloud and MIR green valley early-type galaxies. 
   
Figure \ref{cum_sig5}(c) shows the cumulative distribution
  only for late-type galaxies.
The confidence level to reject the hypothesis that 
  the two distributions
  are extracted from the same parent population
  is 93\% between MIR green valley and 
  MIR blue cloud late-type galaxies, and
  is 97\% between MIR green valley and MIR SF sequence late-type galaxies.
  
Because galaxy properties also strongly depend on stellar mass 
  as well as on environment
  \citep{tasca+09,hwang+10,cucciati+10,bolzonella+10,peng+10,sobral+11},
  we restrict our analysis to only massive galaxies
  with ${\rm log}(M_{\rm star}/M_{\odot})>10.5$ to 
  reduce the mass effect on the MIR properties.
Figure \ref{cum_sig5}(d) shows the cumulative distribution
  for the massive galaxies with 12 $\micron$ detection.
Although the number of galaxies in each MIR class is reduced, 
  the result is similar to that in panel (a). 
Panel (e) shows the case of massive early-type galaxies. 
MIR blue cloud and MIR green valley galaxies again show
  similar distributions.
However, the cumulative distribution of
  MIR green valley galaxies 
  among the massive late-type galaxies (panel f)
  is still between MIR blue cloud and MIR SF sequence galaxies.

\begin{figure}
\centering
\includegraphics[scale=0.65]{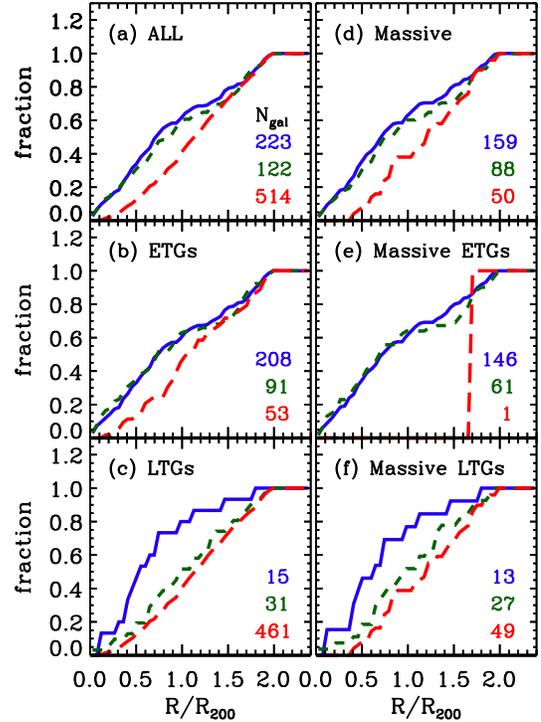}
\caption{Same as Figure \ref{cum_sig5}, but the cumulative distribution as
 a function of cluster/group-centric distance normalized by virial radius ($R/\rvir$).}
\label{cum_rvir}
\end{figure}
  
We perform similar analysis in Figure \ref{cum_rvir}
  using the normalized cluster/group-centric distance $R/\rvir$ 
  instead of $\Sigma_5$.
Panel (a) shows the cumulative distribution of each MIR class
  using all the 12 $\micron$ detected galaxies.  
The $R/\rvir$ distribution of MIR SF sequence galaxies 
  is significantly different from those of MIR blue cloud and 
  MIR green valley galaxies;
  the hypothesis that the two distributions
  are extracted from the same parent population can be
  rejected at a confidence level of $>$99\%.
However, the distributions of
  MIR blue cloud and MIR green valley galaxies are very similar.
They show similar distributions even if we consider only
  early-type galaxies in panel (b),
  massive galaxies in panel (d), and
  massive early-type galaxies in panel (e).

On the other hand,
  when we consider late-type galaxies,
  the similar distribution between MIR blue cloud and MIR green valley galaxies
  disappears.
The confidence level to reject the hypothesis that 
  the two distributions
  are extracted from the same parent population
  is 92\% in panel (c)
  and is 73\% in panel (f).
The cumulative distribution of 
  MIR green valley late-type galaxies appears to be
  similar to that of MIR SF sequence late-type galaxies,
  and be still between the MIR blue cloud and the MIR SF sequence galaxies
  (panels c and f).

Both Figures \ref{cum_sig5} and \ref{cum_rvir} give similar results.
The cumulative distributions of $\Sigma_5$ and $R/\rvir$ 
  for MIR green valley galaxies are between 
  those for MIR blue cloud galaxies and for MIR SF sequence galaxies. 
However, among early-type galaxies the difference between MIR blue cloud 
  and MIR green valley galaxies disappears. 
In contrast, when considering only late-type galaxies,
  the MIR green valley galaxies are still between
  the MIR blue cloud and the MIR SF sequence galaxies.

\section{Discussion}\label{discuss}

\subsection{Star Formation Activity of Galaxy Systems and their Dependence on Virial Mass}

We compare the star formation activity of X-ray bright galaxy groups and clusters
  in the A2199 supercluster using $\Sigma{\rm SFR}/M_{200}$ 
  (\S \ref{isfr}) 
  and the fraction of MIR SF sequence galaxies (\S \ref{mirp}).
Interestingly, both $\Sigma{\rm SFR}/M_{200}$ 
  and the fraction do not seem to depend on virial masses
  of galaxy systems (see Figures \ref{m200_ssfr} and \ref{m200_mirfrac}).
This is seen both in inner ($R\leq0.5\rvir$) 
  and outer regions ($0.5\rvir<R\leq\rvir$). 

Our results are consistent with those in other studies. 
For example, \citet{goto05} used the SDSS data for 115 nearby clusters,
  and found no dependence of $\Sigma{\rm SFR}/M_{200}$
  and of blue/late-type galaxy fraction on cluster virial mass
  (see also \citealt{bai+09,bai+10,balogh+10}).
A recent study based on \textit{WISE} data for 69 clusters at $z<0.1$ 
  also find no correlation between $\Sigma{\rm SFR}/M_{200}$ and
   $M_{200}$ \citep{chung+11}.
   
No correlation between the fraction of passive (or red)
  galaxies and the halo masses of groups/clusters
  could result from a simultaneous increase in the number of member galaxies 
  and of massive quiescent elliptical galaxies 
  with increasing $M_{200}$ of the systems \citep{delucia+12}.
No correlation between the two can also be
  explained by the pre-processing mechanism \citep{zabludoff98},
  which is that giant elliptical galaxies or cD galaxies are formed 
  by galaxy-galaxy mergers in poor groups, and then
  fall into clusters later.
This pre-processing occur in poor groups 
  with $M_{200}<10^{13}~M_{\odot}$. 
This results in 
  no $M_{200}$-dependence of the star formation activity in groups/clusters 
  when the poor groups fall into more massive clusters 
  with $M_{200}>10^{13}~M_{\odot}$ \citep{balogh+10};
  this covers the mass range of the galaxy systems in the A2199 supercluster.
 
On the other hand, some studies show that 
  there is mass dependence
  of the star formation activity of galaxy groups/clusters.
For example, \citet{finn+05} and \citet{koyama+10} 
  found an anti-correlation between 
  $\Sigma{\rm SFR}/M_{200}$ and $M_{200}$ for clusters at $z<1$
  (see also \citealt{bai+07} and \citealt{poggianti+08}).
This anti-correlation is seen in clusters with 
  $10^{13}~M_{\odot}<M_{200}<3\times10^{14}~M_{\odot}$
  that covers our galaxy systems in the A2199 supercluster.
Furthermore, \citet{popesso+12} used far-infrared Herschel data 
  for nine clusters 
  ($M_{200}\geq3\times10^{14}~M_{\odot}$)
  and nine groups/poor clusters 
  ($10^{13}~M_{\odot}<M_{200}<3\times10^{14}~M_{\odot}$) at $z<1.5$, 
  and found that $\Sigma{\rm SFR}/M_{200}$ is higher for groups/poor clusters 
  than for massive clusters at a given redshift. 
They suggested that this mass dependent star formation activity 
  results from earlier quenching of the star formation activity in galaxies 
  associated to more-massive halos (i.e., clusters). 
  
Similarly, some studies show that
  the fraction of SF galaxies decreases with increasing cluster mass
  (e.g., \citealt{martinez+02,rasmussen+12}).
\citet{poggianti+06} used the clusters at low $z$ ($0.04<z<0.08$)
  and at high $z$ ($0.4<z<0.8$), and found an anti-correlation 
  between the fraction of SF galaxies and 
  velocity dispersion ($\sigma_p$) of the clusters;
  the anti-correlation is significant for low-$z$ clusters 
  with $\sigma_p < 550~\kms$ ($M_{200}\lesssim2\times10^{14}M_{\odot}$).
Most galaxy systems in the A2199 supercluster have $\sigma_p < 550~\kms$. 
However, the fraction of SF galaxies (i.e., MIR SF sequence galaxy fraction)
  in galaxy groups/clusters of the A2199 supercluster 
  does not show any strong dependence on $M_{200}$.
  
It seems that the mass dependence of star formation activity of groups/clusters
  is still inconclusive. Some previous studies used heterogeneous data 
  with different wavelength, sky coverage, redshift and sensitivity (see Appendix of \citealt{bai+07}).
  \citet{finn+05} and \citet{koyama+10} suggested that the mass dependence of 
  $\Sigma{\rm SFR}/M_{200}$ appears due to the strong redshift dependence 
  (high-z clusters have higher $\Sigma{\rm SFR}/M_{200}$ and smaller $M_{200}$ 
  in their data). In addition, \citet{poggianti+06} showed that their 
  mass dependence of the fraction of SF galaxies becomes unclear when 
  they used another cluster sample from \citet{miller+05}.
  Thus it still needs more efforts to examine whether cluster mass is a critical factor in 
  determining their star formation activity or not. 
  
If the star formation activity in groups/clusters indeed depends on $M_{200}$, 
  no $M_{200}$-dependence for the galaxy systems in the A2199 supercluster
  indicates that there could be other mechanisms
  that have affected the star formation activity of the systems.
One possible mechanism is the interaction between infall groups and 
  the supercluster. 
If the star formation activity of group galaxies have gradually decreased 
  through the gravitational and/or hydrodynamical
  interaction with the supercluster, 
  the current star formation activity of the groups would depend on 
  when the groups fell into the supercluster environment.
Thus, the different infall time of the groups in the supercluster
  could explain the different star formation activity of groups at a given $M_{200}$. 

Similarly, the interaction between groups/clusters
  can change their star formation activity 
  \citep{bekki99,owen+05,miller05,ferrari+05,johnston+08,hwang+09}. 
In fact, there are three clusters close to 
  the central region of the A2199 supercluster: A2199, A2197E and A2197W.
In particular, A2197E and A2197W overlap each other
  both on the sky and in redshift space; 
  two clusters may be currently interacting. 
A2199 also seems to interact with A2197W/E and 
  with another nearby group, NRGs388.
Thus, the star formation activity of A2199, A2197W/E, and NRGs388 could be enhanced 
  by the interaction among them.
Note that the merging/interaction stages are also important
  to determine the star formation activity of galaxy groups/clusters \citep{hwang+09}.
Therefore,
  the different merging/interaction stages of the groups/clusters
  in the supercluster can add significant scatters to 
  the anti-correlation between the star formation activity and $M_{200}$,
  and can results in no correlation between the two.

\subsection{Morphology Dependence of the Galaxies in the MIR Green Valley}\label{d_envir}

In \S \ref{local} we compare the cumulative distribution 
  of $\Sigma_5$ and $R/\rvir$ for different MIR classes.
The different distributions between
  MIR blue cloud galaxies and MIR SF sequence galaxies
  are consistent with the expectations
  from the morphology- and SFR-density relations;
  most MIR blue cloud galaxies are morphologically early types with
  low SFRs, but MIR SF sequence galaxies are generally late-type galaxies 
  with high SFRs (see Figure \ref{stmass_sfr}).
Interestingly, the distribution of MIR green valley galaxies
  are between
  MIR blue cloud and MIR SF sequence galaxies
  in both environmental indicators.
Because the environment strongly affects galaxy morphology
  and the star formation activity \citep{park+07,bm09},
  the MIR green valley galaxies between MIR blue cloud and 
  MIR SF sequence galaxies seem to be in a transition stage.  
  
However, when we fix galaxy morphology,
  the behavior of the MIR green valley galaxies changes.
The MIR green valley galaxies with early-type morphology
  show a distribution very similar to that of the early-type MIR blue cloud galaxies.
However, the MIR green valley galaxies with late-type morphology
  still show a distribution between MIR blue cloud and 
  MIR SF sequence galaxies.
This trend is also apparent even if we use only 
  massive galaxies with ${\rm log}(M_{\rm star}/M_{\odot})>10.5$.
The similarity between MIR green valley and the MIR blue cloud
  galaxies among the early types is also seen in their stellar masses: 
  ${\rm log}(M_{\rm star}/M_{\odot})=10.66\pm0.04$ for
  MIR green valley early types and 
  ${\rm log}(M_{\rm star}/M_{\odot})=10.75\pm0.03$ for MIR blue cloud
  early types. 
  
\begin{figure}
\centering
\includegraphics[scale=0.65]{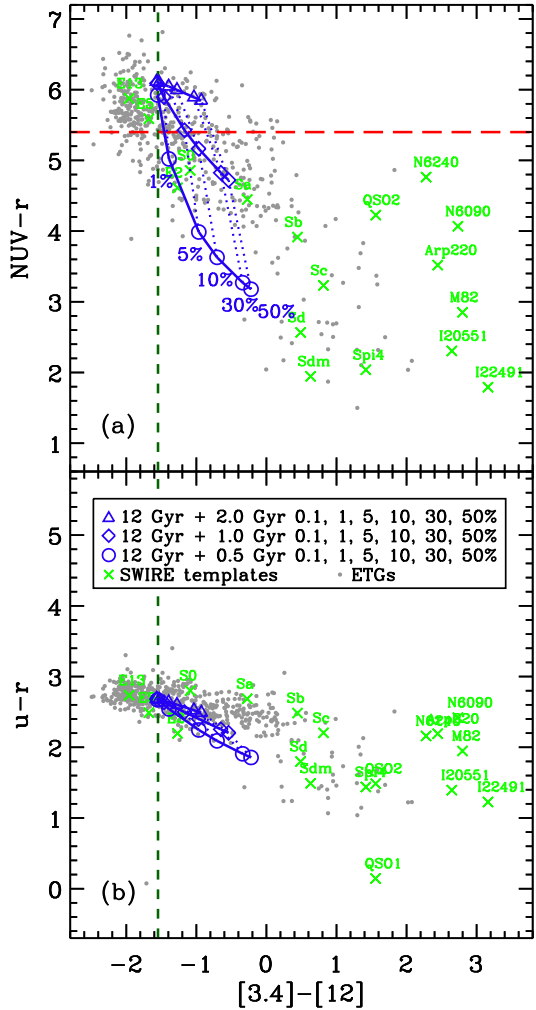}
\caption{(a) NUV$-r$ vs. $[3.4]-[12]$ and (b) $u-r$ vs. $[3.4]-[12]$ 
color-color distributions for early-type member galaxies (gray dots).
Solid lines show the color prediction from the two-component SSP model \citep{piovan+03}. 
These colors are obtained using several combinations of an old stellar population (12 Gyr) and young 
populations (0.5, 1, and 2 Gyr) with varying fraction (0.1, 1, 5, 10, 30, and 50\%) of young populations.
Crosses represent the SWIRE templates \citep{polletta+07}. The vertical dashed line represents 
$[3.4]-[12]=-1.55$ that is the border line between MIR blue cloud and MIR green valley galaxies, 
while the horizontal dashed line is the NUV excess cut, NUV$-r=5.4$, used in \citet{ko+13}.}
\label{mc_ccd}
\end{figure}
  
To better understand the physical meaning of the observed galaxy colors, 
  we show NUV$-r$ and $u-r$ colors of early-type galaxies 
  as a function of \textit{WISE} $[3.4]-[12]$ colors in Figure \ref{mc_ccd} (see also \citet{ko+13}).
  We model that the MIR emission among early types (i.e., MIR green valley early-type galaxies) 
  can be explained by the presence of a small fraction of young stellar population.
  Panel (a) shows the NUV$-r$ versus $[3.4]-[12]$ color-color distribution of
  early-type member galaxies and the SWIRE templates of \citet{polletta+07}.
  We overplot the two-component SSP models of \citet{piovan+03} with
  various combination of an old stellar population (12 Gyr) and
  young populations (0.5, 1, and 2 Gyr) with solar metallicity.
  The model grids indicate different fractions of a young population from 0.1\% to 50\%.

The plot shows that only $>1\%$ of 0.5 Gyr or $>5\%$ of 1 Gyr population 
  can make $[3.4]-[12]$ colors green ($>-1.55$).
  The plot also shows that about 50\% of MIR green valley early-type galaxies 
  have no NUV excess emission (NUV$-r>5.4$). \citet{ko+13} suggested that 
  young stellar populations with 0.5 and 1 Gyr can make the galaxies 
  to have both MIR and NUV excess, but 2 Gyr population does not make any NUV excess.
  This means that half of MIR green valley early-type galaxies stop forming stars 
  at least 2 Gyr ago. This $>2$ Gyr old recent star formation is detected 
  only in the MIR (neither in the NUV nor in the optical) 
  because the MIR traces star formation over longer timescale ($>2$ Gyr) 
  than NUV and optical bands \citep{ko+13,schawinski+14}.

Figure \ref{mc_ccd}(b) displays the $u-r$ versus $[3.4]-[12]$ distribution of 
  early-type member galaxies.
  The model grids show that $u-r$ colors change very little 
  with increasing the fraction of young populations.
  This result is consistent with the results in Figures \ref{opt_CMD} and \ref{stmass_sfr}; 
  the MIR green valley and MIR blue cloud early-type galaxies show similar optical properties.

The environmental dependence of these MIR classes and
 their optical properties shown in \S \ref{mirc}
 suggest a possible evolutionary scenario 
 of galaxies from late-type SF galaxies to early-type quiescent galaxies:
  1) Late-type MIR SF sequence galaxies $\rightarrow$
  2) Late-type MIR green valley galaxies $\rightarrow$
  3) Early-type MIR green valley galaxies $\rightarrow$
  4) Early-type MIR blue cloud galaxies.
  Because the timescale for the suppression of star formation activity 
  is shorter than the timescale for morphological transformation \citep{bm09},
  we expect that the star formation activity of supercluster galaxies decreases first 
  (1$\rightarrow$2),
  and then the morphological transformation would be followed (2$\rightarrow$3).
  \citet{skibba+09} found that the color-density relation is more fundamental 
  than the morphology-density relation. \citet{bamford+09} also obtained a similar result, 
  and suggested that the color transformation driven by environment should take place 
  on shorter timescales than the morphological transformation. 
  This agrees well with our scenario.

\citet{masters+10} studied the environmental dependence of (optically) red 
  (or passive) spiral galaxies that are considered as transition objects 
  between normal blue spirals and red early-type galaxies. 
  They showed that red spirals are mainly found in intermediate density regions 
  and have higher stellar masses than blue spirals. 
  These characteristics of red spirals are very similar to those of 
  late-type MIR green valley galaxies 
  (see Figures \ref{stmass_sfr}, \ref{cum_sig5}, and \ref{cum_rvir}).
  They also suggested that strangulation or starvation 
  (removal of gas in outer halo and no further accretion of cold gas) 
  is a plausible mechanism for producing red spirals.
  These mechanisms can explain quenching of star formation even in low-mass groups 
  \citep{kawata+08,wetzel+13}, 
  which accounts for the presence of late-type MIR green valley located outside the 
  cluster regions (see Figure \ref{galmap_mir}). 
  However, low-mass spirals cannot keep their spiral structures 
  from the environmental processes \citep{bekki+02}. 
  This explains why late-type MIR green valley galaxies (red spirals) 
  have higher stellar masses than late-type MIR SF sequence galaxies (normal blue spirals).

When galaxies enter the MIR green valley, they undergo morphological transformation 
  from late types to early types. 
  Early-type MIR green valley galaxies are more likely to exist in high density regions than 
  late-type MIR green valley galaxies (see \S\ref{local}). 
  This suggests that the environment can play a role in the morphological transformation 
  in the MIR green valley.
  \citet{george+13} suggested that the observed morphological dependence of galaxies 
  on environment cannot be fully explained by strangulation and disk fading \citep{quilis+00}.
  Instead, galaxy mergers and close tidal interactions should be required 
  for the morphological transformation. 
  \citet{park+09} also found that cumulative galaxy-galaxy gravitational/hydrodynamic 
  interactions are the main drive for the morphological transformation in cluster environment. 
  \citet{bekki+11} showed that the transformation in group environment can be driven 
  by repetitive slow encounters with group members. 
  These processes seem suitable for MIR green valley galaxies 
  that are mainly in groups and clusters.

To summarize, the star formation of galaxies is quenched 
  before the galaxies enter the MIR green valley, 
  which is mainly driven by strangulation or starvation. 
  Then, the morphological transformation from late types to early types 
  occurs in the MIR green valley, especially, for massive late-type galaxies 
  with ${\rm log}~(M_{\rm star}/M_{\odot})>10$.
  The main environmental mechanisms for the morphology transformation are 
  galaxy-galaxy mergers and interactions. 
  After the transformation, early-type MIR green valley galaxies stay for several Gyrs 
  keeping the memory of their last star formation.

\section{Summary and Conclusions}\label{conclusions}

Using the multi-wavelength data covering the entire region 
  of the A2199 supercluster,
  we study the star formation activity of galaxy groups/clusters in the supercluster and
  the MIR properties of the supercluster galaxies.  
We determine $\Sigma{\rm SFR}/M_{200}$ of groups and clusters
  using \textit{WISE} 22 $\micron$ data, 
  and find no dependence of $\Sigma{\rm SFR}/M_{200}$ on $M_{200}$.
  
We classify galaxies into three MIR groups 
  in the \textit{WISE} $[3.4]-[12]$ color versus   
  12 $\micron$ luminosity diagram: 
  MIR blue cloud galaxies, MIR SF sequence galaxies, 
  and MIR green valley galaxies. 
MIR blue cloud galaxies are mostly (90\%) early types,
  while MIR SF sequence galaxies are predominantly (93\%) late types.
MIR green valley galaxies consist of  early- (68\%)
  and late-type galaxies (32\%).
SFRs for MIR blue cloud and MIR green valley galaxies are similar, 
  but much smaller than those for MIR SF sequence galaxies.
  It is important that the MIR green valley galaxies are distinguishable 
  from the optical green valley galaxies, in the sense that they belong to
  the optical red sequence. Thus star formation quenching of galaxies occurs 
  before the galaxies enter the MIR green valley.
 
We calculate the fraction of each MIR class in groups/clusters.
We could not find any dependence of the fraction
  of each MIR class on $M_{200}$,
  consistent with the trend of $\Sigma{\rm SFR}/M_{200}$.
These results suggest that group/cluster mass does not play 
  an important role in controlling the global star formation activity of the systems.
On the other hand,
  there could be other mechanisms affecting the global star formation activity of the systems.
These include the interaction between infalling groups and the supercluster,
  and between groups/clusters.
     
We compare the cumulative distribution of $\Sigma_5$ and $R/\rvir$ 
  for the three MIR classes. 
MIR green valley galaxies show the distribution between
  MIR blue cloud and MIR SF sequence galaxies.
When only considering early-type galaxies, 
  the difference between 
  MIR blue cloud and MIR green valley galaxies disappears. 
However, late-type galaxies in the MIR green valley
  still show the distribution between
  MIR blue cloud and MIR green valley galaxies.
These results suggest a possible evolutionary scenario:
1) Late-type MIR SF sequence galaxies $\rightarrow$
2) Late-type MIR green valley galaxies $\rightarrow$
3) Early-type MIR green valley galaxies $\rightarrow$
4) Early-type MIR blue cloud galaxies.
Thus the MIR green valley is the site 
where morphology transformation of galaxies mainly appears to occur.

\acknowledgments
We thank Margaret Geller for many helpful discussions. 
G.H.L. acknowledges the support by the National Research Foundation of Korea (NRF) Grant funded 
by the Korean Government (NRF-2012-Fostering Core Leaders of the Future Basic Science Program).
M.G.L was supported by the NRF grant funded by the Korea Government (MEST) (No. 2012R1A4A1028713).
H.S.H. acknowledges the Smithsonian Institution for the support of his post-doctoral fellowship.
J.S. was supported by Global Ph.D. Fellowship Program through an NRF funded by the MEST (No.2011-
0007215).
H. Shim is supported by Basic Science Research Program through the National Research Foundation of Korea (NRF) 
funded by the Ministry of Science, ICT \& Future Planning (2014R1A1A1038088). 
AD acknowledges partial support from the INFN grant In-Dark and from the grant Progetti di Ateneo/CSP TO Call2 2012 0011 ``Marco Polo'' of the University of Torino.
This publication makes use of data products from the {\it Wide-field Infrared Survey Explorer},
which is a joint project of the University of California, Los Angeles,
and the Jet Propulsion Laboratory/California Institute of Technology,
funded by the National Aeronautics and Space Administration.
Funding for the SDSS and SDSS-II has been provided by the Alfred P. Sloan
Foundation, the Participating Institutions, the National Science
Foundation, the U.S. Department of Energy, the National Aeronautics and
Space Administration, the Japanese Monbukagakusho, the Max Planck
Society, and the Higher Education Funding Council for England.
The SDSS Web site is http://www.sdss.org/.
The SDSS is managed by the Astrophysical Research Consortium for the
Participating Institutions. The Participating Institutions are the
American Museum of Natural History, Astrophysical Institute Potsdam,
University of Basel, Cambridge University, Case Western Reserve University,
University of Chicago, Drexel University, Fermilab, the Institute for
Advanced Study, the Japan Participation Group, Johns Hopkins University,
the Joint Institute for Nuclear Astrophysics, the Kavli Institute for
Particle Astrophysics and Cosmology, the Korean Scientist Group, the
Chinese Academy of Sciences (LAMOST), Los Alamos National Laboratory,
the Max-Planck-Institute for Astronomy (MPIA), the Max-Planck-Institute
for Astrophysics (MPA), New Mexico State University, Ohio State University,
University of Pittsburgh, University of Portsmouth, Princeton University,
the United States Naval Observatory, and the University of Washington.

\end{document}